\documentclass[usenatbib,useAMS,manuscript]{mn2e}

\usepackage{CJK}

\usepackage{amsmath,amsfonts,amssymb,bm}
\usepackage{graphicx,graphics,tabularx,natbib,color,multicol,pdflscape}
\usepackage{siunitx,threeparttable,pdflscape}

\begin{document}

\title{\textbf{Halo Acceleration Relation}}

\author[Tian \& Ko]
{Yong Tian$^{1}$\thanks{E-mail:yongtian@astro.ncu.edu.tw}
and Chung-Ming Ko$^{1,2}$\thanks{E-mail:cmko@astro.ncu.edu.tw}
\\
$^{1}$Institute of Astronomy, National Central University, Taoyuan City, Taiwan 32001, Republic of China \\
$^{2}$Department of Physics and Centre for Complex Systems, National Central University, Taoyuan City, Taiwan 32001, Republic of China
}

\date{Accepted 2019 May 31. Received 2019 May 28; in original form 2019 April 03}

\maketitle

    \begin{abstract}
    Recently, from the new Spitzer Photometry and Accurate Rotation Curves (SPARC) data, McGaugh et al. (2016) reported a tight Radial Acceleration Relation (RAR) between the observed total acceleration and the acceleration produced by baryons in spiral galaxies.
    The relation can be fitted by different functions.
    However, these functions can be discerned if we express the data in the form of halo acceleration relation (HAR). The data reveals a maximum in the halo acceleration.
    We examined NFW (cusp) and Burkert (core) profiles in the context of dark matter and different parameter families of the interpolating function in the framework of Modified Newtonian Dynamics (MOND).
    \end{abstract}

\begin{keywords}
gravitation -- gravitational lensing: strong -- galaxies: elliptical and lenticular, cD -- galaxies: kinematics and dynamics -- dark matter
\end{keywords}

    \section{Introduction}\label{sec:intro}
    Flat rotational curve in spiral galaxies exemplified baryonic mass is not enough to explain the needed dynamical mass under Newtonian dynamics \citep{Bosma81, Rubin82}.
    This is the famous missing mass problem.
    One may ask is the mass discrepancy related to any (observable) physical quantity?
    When testing the prediction of Modified Newtonian Dynamics (MOND) \citep{Milgrom83}, \cite{Sanders90} revealed that the mass discrepancy is related to gravitational acceleration.
    With the data from 60 spiral galaxies, \citet{McGaugh04} showed that mass discrepancy and gravitational acceleration is indeed related.
    The relation is called the Mass Discrepancy-Acceleration Relation (MDAR).

    The ratio of the dynamical mass to the baryonic mass increases as acceleration decreases beyond the scale $\mathfrak{a}_0=1.2\times10^{-10}$ m\,s$^{-2}$.
    If we take gravitational acceleration as $g(r)=GM(<r)/r^2$ at radius $r$ from the centre, the mass ratio is the same as the acceleration ratio.
    MDAR can be interpreted as acceleration discrepancy relation.
    Furthermore, in the framework of dark matter, discrepancy in mass can be considered as the halo mass and the acceleration discrepancy as the halo acceleration.

    Using high precision data from 153 spiral galaxies in SPARC (Spitzer Photometry and Accurate Rotation Curves) database,
    \citet{McGaugh16} obtained a tight Radial Acceleration Relation (RAR) of observed acceleration ($g_{\rm o}$)
    and baryonic acceleration ($g_{\rm b}$), see left panel of Fig.~\ref{fig:MHA}.
    Roughly, RAR can be described approximately by a broken power law.
    The asymptotic relation in small acceleration regime is $g_{\rm o}=\sqrt{\mathfrak{a}_0g_{\rm b}\,}$,
    and in the large acceleration regime $g_{\rm o}=g_{\rm b}$.
    The turning point is somewhere around $\mathfrak{a}_0$.
    The relation can be written as
    \begin{equation}\label{eq:RAR}
      g_{\rm o}=\nu(y)g_{\rm b}\,,
    \end{equation}
    where $y=g_{\rm b}/\mathfrak{a}_0$, and $\nu(y)$ (called the interpolating function) behaves asymptotically:
    $\nu(y)\simeq 1$ as $y\gg 1$ and $\nu(y)\approx y^{-1/2}$ as $y\ll 1$.

    \citet{McGaugh16} suggested a form first proposed by \cite{MS08}
    \begin{equation}\label{eq:nuform}
      \nu_{\rm MS}(y)=(1-e^{-\sqrt{y}})^{-1}\,.
    \end{equation}
    Consistent results from the dynamics in elliptical galaxies were found \citep{Scarpa06, Janz16, TK16, Lelli17}.
    \cite{Milgrom19} even extended it to galaxy groups down to $10^{-14}$\,m\,s$^{-2}$ for baryonic acceleration.
    Moreover, the relation also holds in relativistic effect such as gravitational lensing \citep[][showed the relation in 57 Einstein rings]{TK17}.

    The origin of the RAR raises different possibilities for dark matter model and modified gravity \citep{Lelli17}.
    \cite{WK15} analyzed cosmological simulation of galaxy formation in dark matter framework, and found poor agreement with the data in  \cite{McGaugh04}.
    Studied from the EAGLE hydrodynamic simulation, \cite{Ludlow17} showed that the RAR is a result of galaxy formation with a larger acceleration scale ($g_\dagger=2.6\times10^{-10}$\,m\,s$^{-2}$).
    On the other hand, \cite{Li18} performed a Bayesian analysis on individual galaxies in SPARC database.
    They found a very small scatter in acceleration scale $\mathfrak{a}_0=1.2\pm0.02\times10^{-10}$\,m\,s$^{-2}$.
    This poses a serious challenge to dark matter simulation as $g_\dagger$ is 70\,$\sigma$ from the observed value.

    Adopting an abundance-matching prescription \citep{Kravtsov18}, \cite{DL16} was successful to use semi-empirical approach of dark matter model to reproduce the RAR.
    Nevertheless, the intrinsic scatter is considerably large due to uncertainties in baryonic-to-halo mass ratio, the mass-concentration relation of DM halos, the mass-size relation of stellar disks, etc.
	\cite{Desmond17} also got very large scatter even applying zero intrinsic scatter on abundance-matching prescription.
	On the other hand, \cite{Navarro17} claimed that the RAR can be explained by semi-analytical model in $\Lambda$CDM by adopting the
	abundance-matching relation by \cite{Behroozi13}.

    Another possibility is to treat the RAR as the result of modified dynamical law.
    \cite{Milgrom83} proposed MOND to explain the missing mass problem.
    In its original form, MOND is expressed as $\mu(x)g_{\rm o}=g_{\rm b}$,
    where $x=g_{\rm o}/\mathfrak{a}_0$ and $\mu(x)$ behaves asymptotically: $\mu(x)\simeq1$ as $x\gg 1$ and $\mu(x)\approx x$ as $x\ll 1$.
    With $\nu(y)$ as the functional inverse of $\mu(x)$, we get Eq.~(\ref{eq:RAR}).
    Thus RAR is a natural consequence of MOND, and MOND can explain the small intrinsic scatter of the RAR.
    However, RAR can be fitted by many different functional forms of $\nu(y)$ satisfying the asymptotic behaviour (see left panel of Fig.~\ref{fig:MHA}).

    In any case, the observed RAR gives a constraint on $\mu(x)$ and $\nu(y)$,
    \begin{equation}\label{eq:MOND1}
      \frac{d\ln\mu}{d\ln\,x} > 0\,, \quad{\rm and}\quad\frac{d\ln\nu}{d\ln\,y} < 0\,.
    \end{equation}

    If $\mu(x)$ approaches 1 slower than $x^{-1}$ as $x\gg 1$, the halo acceleration has a maximum \citep{BM99}.
    \citet{MS05} showed that the Burkert profile \citep{Burkert95} has a maximum acceleration at the scale radius $r_0$.
    \citet{Donato09} and \citet{Gentile09} revealed the universal maximum halo and baryonic acceleration respectively.
    Especially, the universal maximum baryonic acceleration spans 15 order of B-band absolute magnitude.
    MOND can explain this universal maximum acceleration with the acceleration scale $\mathfrak{a}_0$ \citep{Milgrom09, Milgrom16}.

    In this work, we point out that the data of halo acceleration clearly has one maximum with respect to baryonic acceleration $g_{\rm b}$.
    Moreover, the HAR can distinguish different forms of interpolating functions in MOND or can constrain dark matter profiles.

    \section{Data}\label{sec:Data}

        \begin{table}
        \sisetup{table-number-alignment = center,
                 table-figures-integer = 2,
                 table-figures-decimal = 3}
             \centering
             \setlength{\tabcolsep}{8pt}
             \setlength{\extrarowheight}{1pt}
            \caption[]{\textbf{The binned data of the RAR and HAR of spiral and elliptical
            galaxies.}}\label{tab:Result}
            \begin{tabular}{
            S
            S[separate-uncertainty, table-figures-uncertainty=1]
            S[separate-uncertainty, table-figures-uncertainty=1]
            c}
            \hline\hline
            {$\log_{10}[g_{\rm b}]$} & {$\log_{10}[g_{\rm o}]$} & {$\log_{10}[g_{\rm h}]$} & number \\
            {${\rm m\,s}^{-2}$} & {${\rm m\,s}^{-2}$} & {${\rm m\,s}^{-2}$} & \\
            \hline
             -8.859	&	 -8.848(204)	& -10.416(204)	&	180	\\
             -9.208	&	 -9.175(138)	& -10.304(138)	&	145	\\
             -9.436	&	 -9.337(110)	& -10.028(110)	&	154	\\
             -9.639	&	 -9.509(132)	& -10.096(132)	&	172	\\
             -9.837	&	 -9.668(147)	& -10.161(147)	&	209	\\
            -10.036	&	 -9.792(131)	& -10.158(131)	&	238	\\
            -10.231	&	 -9.933(117)	& -10.236(117)	&	229	\\
            -10.430	&	-10.049(140)	& -10.283(140)	&	242	\\
            -10.642	&	-10.181(120)	& -10.365(120)	&	229	\\
            -10.852	&	-10.322(131)	& -10.474(131)	&	283	\\
            -11.064	&	-10.438(145)	& -10.556(145)	&	323	\\
            -11.273	&	-10.582(148)	& -10.681(148)	&	308	\\
            -11.487	&	-10.706(154)	& -10.784(154)	&	191	\\
            -11.765	&	-10.798(190)	& -10.848(190)	&	75	\\
            \hline\hline
            \end{tabular}
        \end{table}

    We are interested in the observed acceleration $g_{\rm o}$ (or total acceleration), baryonic acceleration $g_{\rm b}$ (acceleration from baryons only)
    and halo acceleration $g_{\rm h}=g_{\rm o}-g_{\rm b}$ (the difference between total and baryonic acceleration).
    In this study, our samples include both spiral and elliptical galaxies.
    The observed acceleration for spirals is obtained by dynamics \citep{McGaugh16} from SPARC database \citep{Lelli16b}, and for ellipticals by gravitational lensing \citep{TK17} from SLACS database \citep{Auger09}.
    The baryonic acceleration comes from total baryonic mass via population synthesis and mass model of galaxies \citep{TK17}.
    Altogether, our sample comprises: \\
    i. 2693 data of 153 spiral galaxies \citep{McGaugh16};\\
    ii. 285 data of 53 elliptical galaxies \citep{TK17}.

    In Fig.~\ref{fig:MHA}, we showed the observed acceleration and the halo acceleration against the baryonic acceleration.
    In the figure, We bin the data in $g_{\rm b}$ of RAR. The binning size depends on the number of data.
    The distribution of data in each is consistent with normal distribution.
    Thus, we take the normal procedure to get the mean and standard deviation.
    The 14 red circles are binned data of spirals only \citep{McGaugh16},
    while the 14 black circles are binned data including spirals and ellipticals, see Table~\ref{tab:Result}.

    Data of elliptical galaxies (all from lensing) is about 10\% of the total.
    However, they are mainly in the high acceleration regime from $10^{-9}$ to $10^{-10}$ m\,s$^{-2}$,
    since strong lensing mostly occurs around $\sim10\,\mathfrak{a}_{0}$ where little dark matter is expected within Einstein rings radius
    \citep[see e.g., explanation in][]{Sanders14}.
    In the range $-8.5\ge \log_{10}g_{\rm b}\ge -10.5$, there are 1254 data points of spirals and 285 ellipticals (about 19\% of total).

    As the data of elliptical galaxies is important for the high acceleration regime of HAR, we need to be cautious in mass estimation.
    The baryonic mass of SLACS lenses is estimated from population synthesis with Salpeter IMF (e.g., \cite{Auger09, TK17}).
    We apply $20\%$ uncertainty in the stellar mass by Salpeter IMF to gauge the influence on the results.
    We find that the effect is significant for halo acceleration $g_{\rm h}$ in the last five binned data at the high acceleration regime.
    But the effect on baryonic acceleration $g_{\rm b}$ is small.
    The discrepancy is less than 0.037 dex for all $g_{\rm b}$.
    However, the effect on halo acceleration $g_{\rm h}$ increases as $g_{\rm b}$ increases.
    We list the uncertainty for the five high acceleration data points in Table \ref{tab:uncertainty}.

        \begin{table}
             \centering
             \setlength{\tabcolsep}{12pt}
             \setlength{\extrarowheight}{4pt}
            \caption[]{\textbf{The last five binned data with $20\%$ uncertainty in the mass deduced from Salpeter IMF}}\label{tab:uncertainty}
            \begin{tabular}{ccc}
            \hline\hline
            {$\log_{10}[g_{\rm b}]$} & {$\log_{10}[g_{\rm o}]$} & {$\log_{10}[g_{\rm h}]$}  \\
            {${\rm m\,s}^{-2}$} & {${\rm m\,s}^{-2}$} & {${\rm m\,s}^{-2}$} \\
            \hline
            -8.859$^{+0.035}_{-0.037}$	& -8.848$^{-0.002}_{-0.005}$	& -10.416$^{+0.505}_{-0.544}$ \\
            -9.208$^{+0.020}_{-0.024}$	& -9.175$^{+0.009}_{-0.001}$	& -10.304$^{+0.182}_{-0.220}$ \\
            -9.436$^{+0.018}_{-0.022}$	& -9.337$^{-0.015}_{-0.000}$	& -10.028$^{+0.064}_{-0.074}$ \\
            -9.639$^{+0.018}_{-0.020}$	& -9.509$^{+0.003}_{+0.006}$	& -10.096$^{+0.044}_{-0.075}$ \\
            -9.837$^{+0.016}_{-0.015}$	& -9.668$^{+0.004}_{+0.007}$	& -10.161$^{+0.066}_{-0.055}$ \\
            \hline\hline
            \end{tabular}
        \end{table}

    \section{Results}\label{sec:Result}

        \begin{figure*}
        \centering
        \includegraphics[width=\columnwidth]{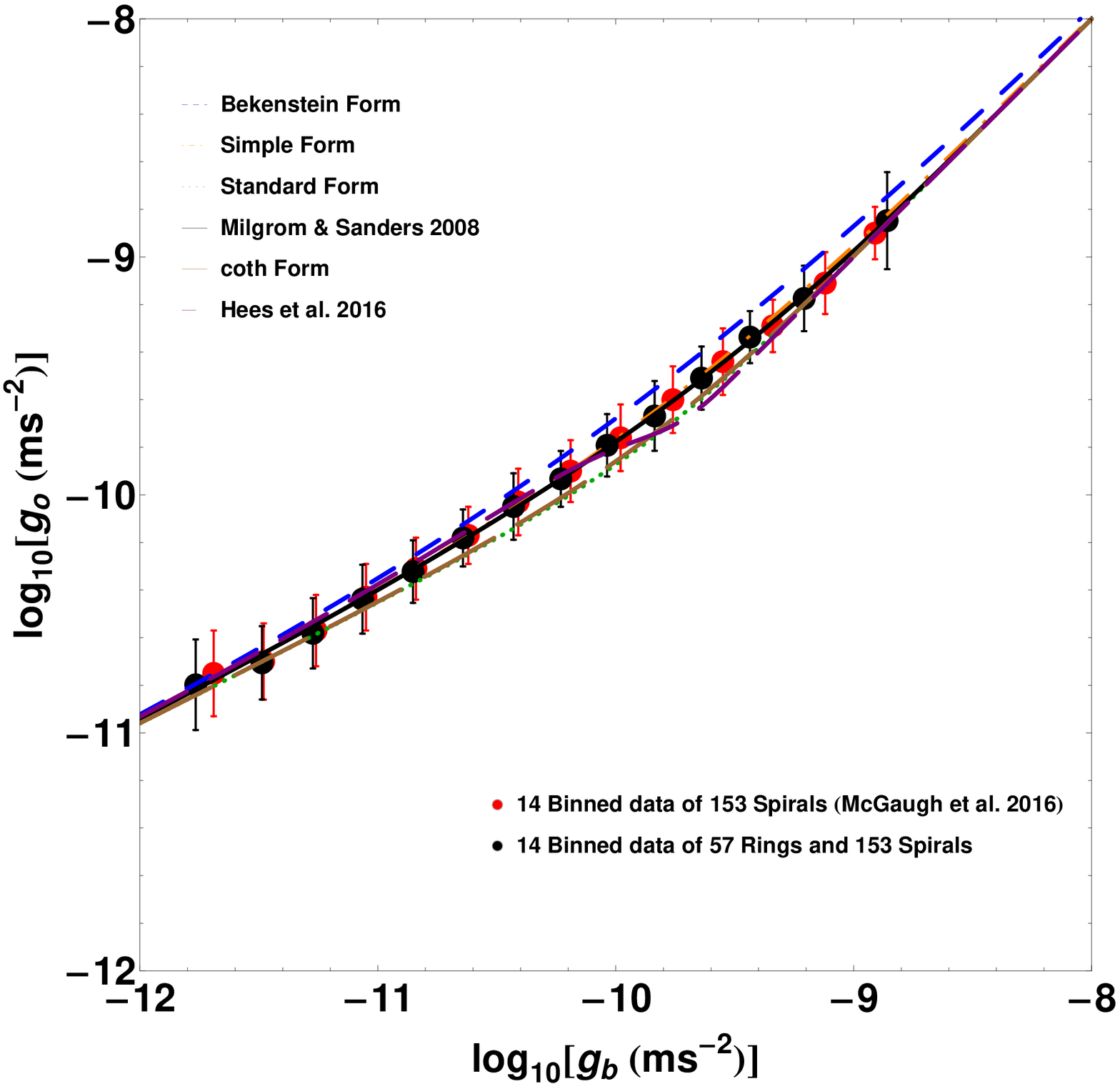}
        \includegraphics[width=\columnwidth]{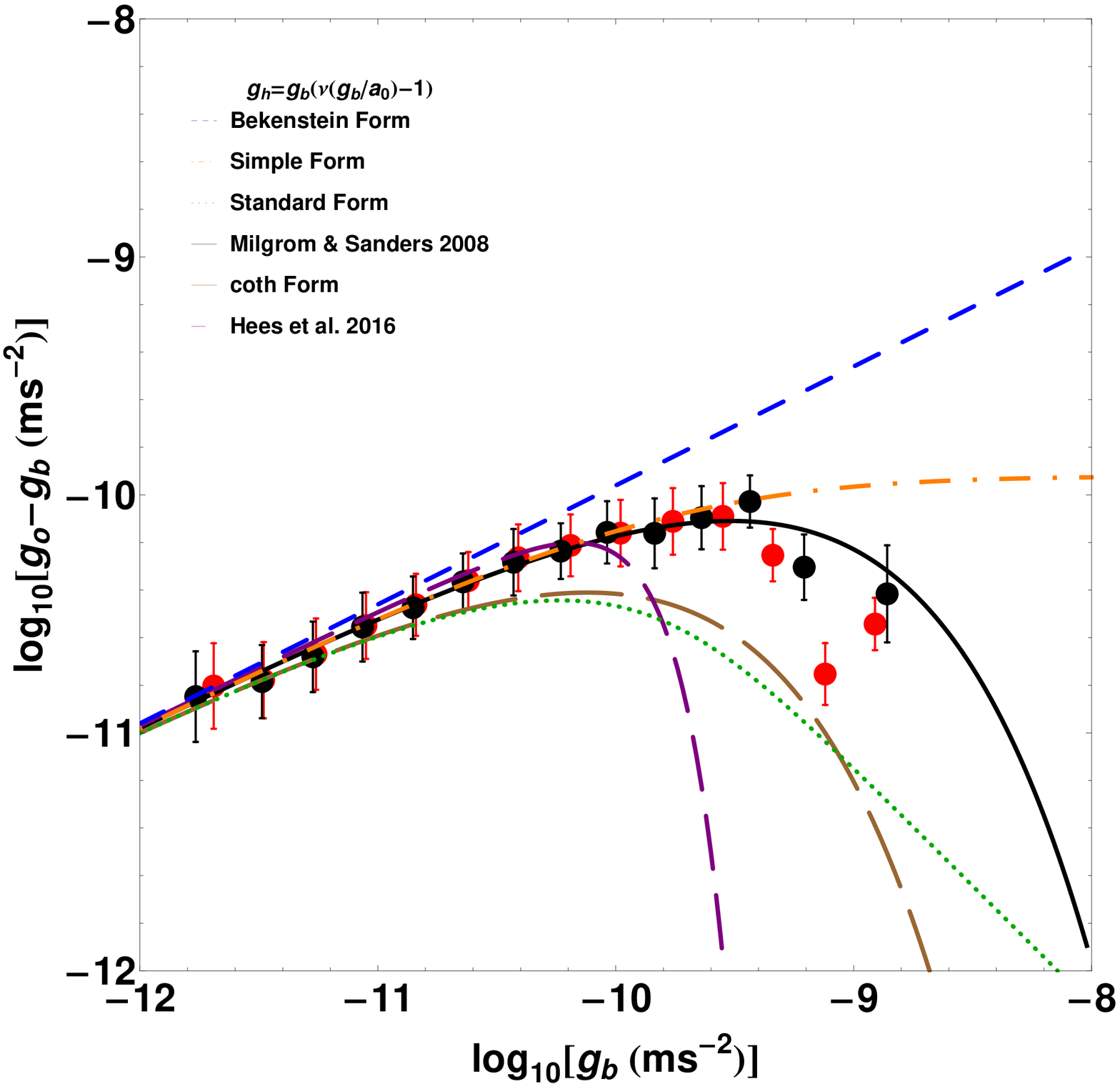}
        \caption{
        The relations between total acceleration and halo acceleration against baryonic acceleration.
        Left panel: Radial acceleration relation (RAR). Right panel: Halo acceleration relation (HAR).
        14 red circles denote binned data of spirals \citep{McGaugh16} and
        14 black circles are binned data including spirals and ellipticals \citep{TK17}.
        Six interpolating functions are plotted for comparison: blue dash, orange dot-dashed, dotted, black thick, brown and purple long dash lines
        representing Bekenstein, simple, standard, MS, coth and Hees forms, respectively (see text for their expressions).
        The HAR can easily distinguish different interpolating functions while RAR cannot.
        }\label{fig:MHA}
        \end{figure*}

        From RAR or Eq.~(\ref{eq:RAR}), we learn that in the high acceleration regime
        (i.e., $g_{\rm b}\gg \mathfrak{a}_0$)
        $g_{\rm o}\approx g_{\rm b}\gg \mathfrak{a}_0$
        (as the interpolating function $\nu(y)\approx 1$ for $y\gg 1$ and $y=g_{\rm b}/\mathfrak{a}_0$).
        Thus, we know that when $g_{\rm b}\gg \mathfrak{a}_0$, $g_{\rm h}=g_{\rm o}-g_{\rm b}\ll g_{\rm b}$, but
        the intriguing question will be how the value of $g_{\rm h}$ compares with $\mathfrak{a}_0$ in this limit.
        Theoretically speaking, depending on the functional form of the interpolating function $\nu(y)$,
        it is possible to have $g_{\rm h}\gg \mathfrak{a}_0$, $\approx \mathfrak{a}_0$ or
        $\ll \mathfrak{a}_0$ when $g_{\rm b}\gg \mathfrak{a}_0$.

        To illustrate idea, let us consider the $\alpha\eta$-form \citep{Chiu11},
        \begin{equation}\label{eq:alphaeta}
          \nu(y)=\left[1+\frac{1}{2}\left(\sqrt{4y^{-\alpha}+\eta^2\,}-\eta\right)\right]^{1/\alpha}\,,
        \end{equation}
        with $\alpha>0$ and $\eta\ge 0$. In fact, this form includes Bekenstein form $(\alpha,\eta)=(1,0)$ \citep{Bekenstein04},
        simple form $(\alpha,\eta)=(1,1)$ \citep{Milgrom86}, and standard form $(\alpha,\eta)=(2,1)$.

        For the case $\eta=0$, $g_{\rm h}=\left(\nu(y)-1\right) g_{\rm b}\approx \mathfrak{a}_0 y^{(1-\alpha/2)}/\alpha$
        as $y\gg 1$ (or $g_{\rm b}\gg\mathfrak{a}_0$), and
        for the case $\eta>0$, $g_{\rm h}\approx \mathfrak{a}_0 y^{(1-\alpha)}/\alpha\eta$ as $y\gg 1$.
        Therefore, when $g_{\rm b}\gg\mathfrak{a}_0$, (1) $g_{\rm h}\approx \sqrt{g_{\rm b}\mathfrak{a}_0\,}\gg \mathfrak{a}_0$ for Bekenstein form;
        (2) $g_{\rm h}\approx \mathfrak{a}_0$ for simple form; and
        (3) $g_{\rm h}\approx \mathfrak{a}_0^2/2g_{\rm b}\ll \mathfrak{a}_0$ for standard form.

        We show our data in Fig.~\ref{fig:MHA}. In the left panel, we plot $g_{\rm o}$ against $g_{\rm b}$ (i.e., the common radial acceleration relation, RAR),
        and in the right panel, we plot $g_{\rm h}=g_{\rm o}-g_{\rm b}$ against $g_{\rm b}$ (we call this the halo acceleration relation, HAR).
        In addition to the three forms discussed above (Bekenstein, simple and standard), we also show three more forms for comparison:
        (4) MS (Milgrom \& Sanders) form (Eq.~(\ref{eq:nuform})) was suggested to fit the RAR of spiral galaxies \citep{McGaugh16};
        (5) Coth form ($\nu(y)=\coth(\sqrt{y})$) is similar to MS form; and
        (6) Hees form ($\nu(y)=[1-\exp(-y^2)]^{-1/4}+3/4\exp(-y^2)$)
        was proposed to fit the data of rotational curve of spiral galaxies and solar system \citep{Hees16}.
        The halo acceleration $g_{\rm h}$ of standard, MS, Coth and Hees forms all approach zero in high acceleration regime
        ($g_{\rm b}\gg \mathfrak{a}_0$).

        In the left panel of Fig.~\ref{fig:MHA}, all six forms of the interpolating function are within the error of either the red circles (binned data for spirals only) or the black circles (binned data of both spirals and ellipticals).
        Thus all forms are a good fit to RAR.
        When $g_{\rm b}$ approaches zero, $g_{\rm o}=\nu(y)g_{\rm b}\rightarrow \sqrt{g_{\rm b}\mathfrak{a}_0\,}$ (as $y\rightarrow 0$).
        Therefore, $\mathfrak{a}_0$ can be determined from the intercept of the figure, which is $\mathfrak{a}_0=1.2\times 10^{-10}$ m\,s$^{-2}$
        \citep[consistent with][]{McGaugh16}.

        In the right panel, the HAR shows a clear unimodal profile.
        $g_{\rm h}$ has a maximum about $0.78\,\mathfrak{a}_0=9.36\times 10^{-11}$ m\,s$^{-2}$, around $g_{\rm b}\approx 3.05\mathfrak{a}_0=3.66\times 10^{-10}$ m\,s$^{-2}$.
        We should point out that these values are uncertain due to the quality of the data in the high acceleration regime.
        Although four forms (standard, MS, Coth and Hees) have unimodal profile, only MS form gives the right position and value of the maximum.
        Recall that the red circles come from spirals data only, and the black circles include both spirals and ellipticals data.
        Moreover, the ellipticals data are mostly in the large $g_{\rm b}$ range.
        It is interesting to note that MS form fits the data better when more high acceleration data are included.

        Recall that RAR gives a constraint on the interpolating function $d\ln\nu/d\ln y$ (Eq.~(\ref{eq:MOND1})).
        Now the existence of a maximum in HAR gives more constraints on the interpolating function: (i) $d\ln(\nu-1)/d\ln y+1=0$ has one root; and
        (ii) $d^2\ln(\nu-1)/d(\ln y)^2<0$.

    \section{Discussion}\label{sec:Dis}

        \begin{figure}
        \centering
        \includegraphics[width=\columnwidth]{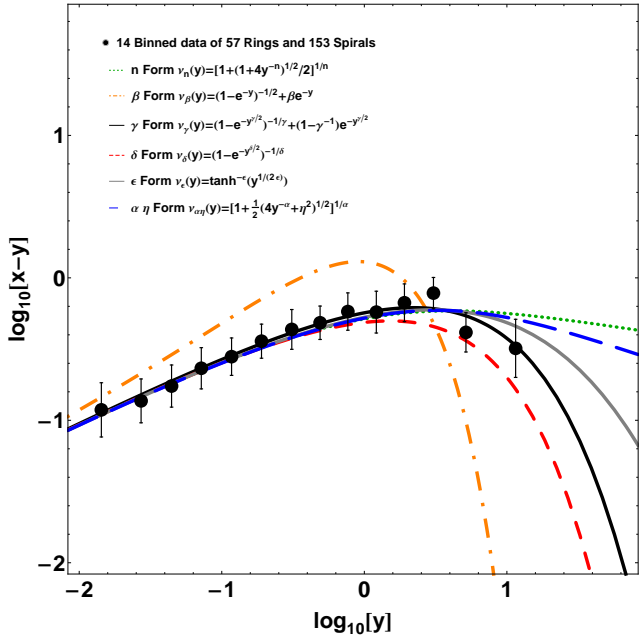}
        \caption{Fitting of the HAR by different interpolating functions.The best fit parameters of the forms are listed in Table 1.
        }\label{fig:Interpolating}
        \end{figure}

        \begin{table}
             \centering
             \setlength\extrarowheight{2.5pt}
            \caption[]{\textbf{Best Fit Parameters of different interpolating functions of MOND.}}\label{tab:Fit}
            \begin{tabular}{ccccccc}
            \hline
            &   $n$  & $\beta$ & $\gamma$ & $\delta$ & $\epsilon$ & $(\alpha,\eta)$ \\
            \hline
            Best Fit & 1.16  & 2.82   & 1.04    &  1.16  & 1.56 & (1.34, 0.57)\\
            $\chi^2$ & 0.37  & 19.7   & 0.15    &  0.37  & 0.27 & 0.32   \\
            $\nu(1)$ & 1.51  & 2.30   & 1.57    &  1.49   & 1.53  & 1.52    \\
            \hline
            \end{tabular}
        \end{table}

        As shown in Fig.~\ref{fig:MHA}, RAR can be fitted by a number of forms of the interpolating function, but HAR is sensitivity to different forms
        especially in the high acceleration regime.
        In literature RAR has been studied under dark matter \citep{WK15, DL16, Desmond17, Ludlow17, Navarro17} and MOND \citep{Milgrom16, Lelli17}.
        Both frameworks are able to come up with an explanation for RAR.
        Here we test both against HAR.

    \subsection{MOND Interpolating Functions}

    We pick six commonly used interpolating functions:
    $n$-form \citep{Kent87}, $\beta$-form \citep{Begeman91}, $\gamma$-form \citep{MS08}, $\delta$-form \citep{McGaugh08}, $\epsilon$-form \citep{Milgrom16}, and $\alpha\eta$-form \citep{Chiu11,TK16,TK17}.
    We perform reduced $\chi^2$ method with our 14 binned data and error bar to fit these forms to HAR, see Fig.~\ref{fig:Interpolating}.
    In the figure, we normalize all the acceleration to
    $\mathfrak{a}_0=1.2\times 10^{-10}$ m\,s$^{-2}$, i.e., $g_{\rm b}/\mathfrak{a}_0=y$, $g_{\rm h}/\mathfrak{a}_0=y\nu(y)-y=x-y$.
    The best fit values of the parameters and reduced $\chi^2$ for all the forms are listed in Table~\ref{tab:Fit}.
    $\gamma$-form is the best among all others as $\gamma=1.041$ according to reduced $\chi^2$.
    In fact, that is basically MS form $\nu_{\rm MS}(y)$, Eq.~(\ref{eq:nuform}).
    We also list the value of $\nu(1)$ for each family with best fit value because \cite{Milgrom16} pointed out $\nu(1)\approx1.6$ would be a good description of the MDAR.

    \subsection{Dark Matter Profile}
    \begin{figure*}
        \includegraphics[width=0.9\columnwidth]{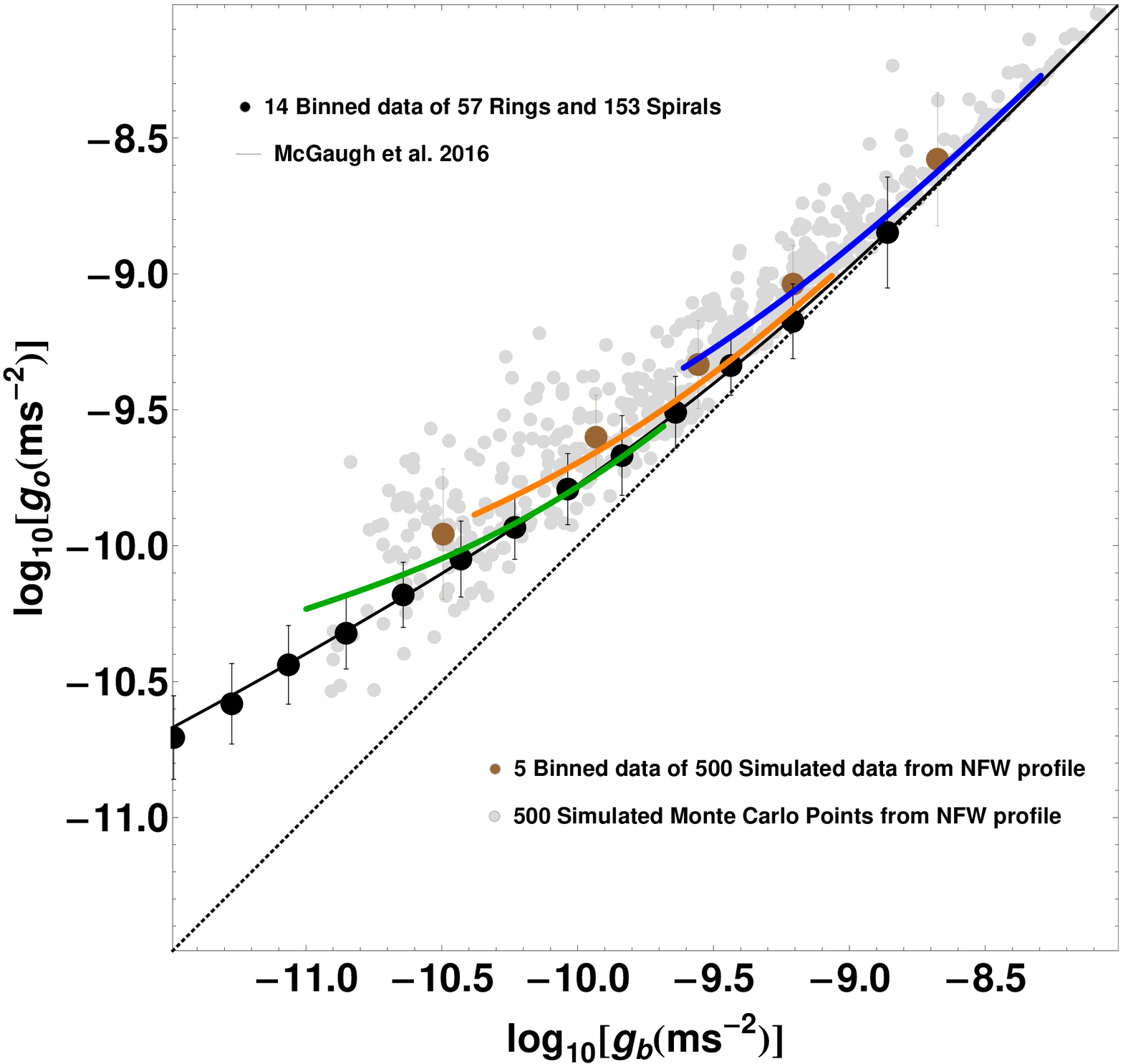}
        \includegraphics[width=0.9\columnwidth]{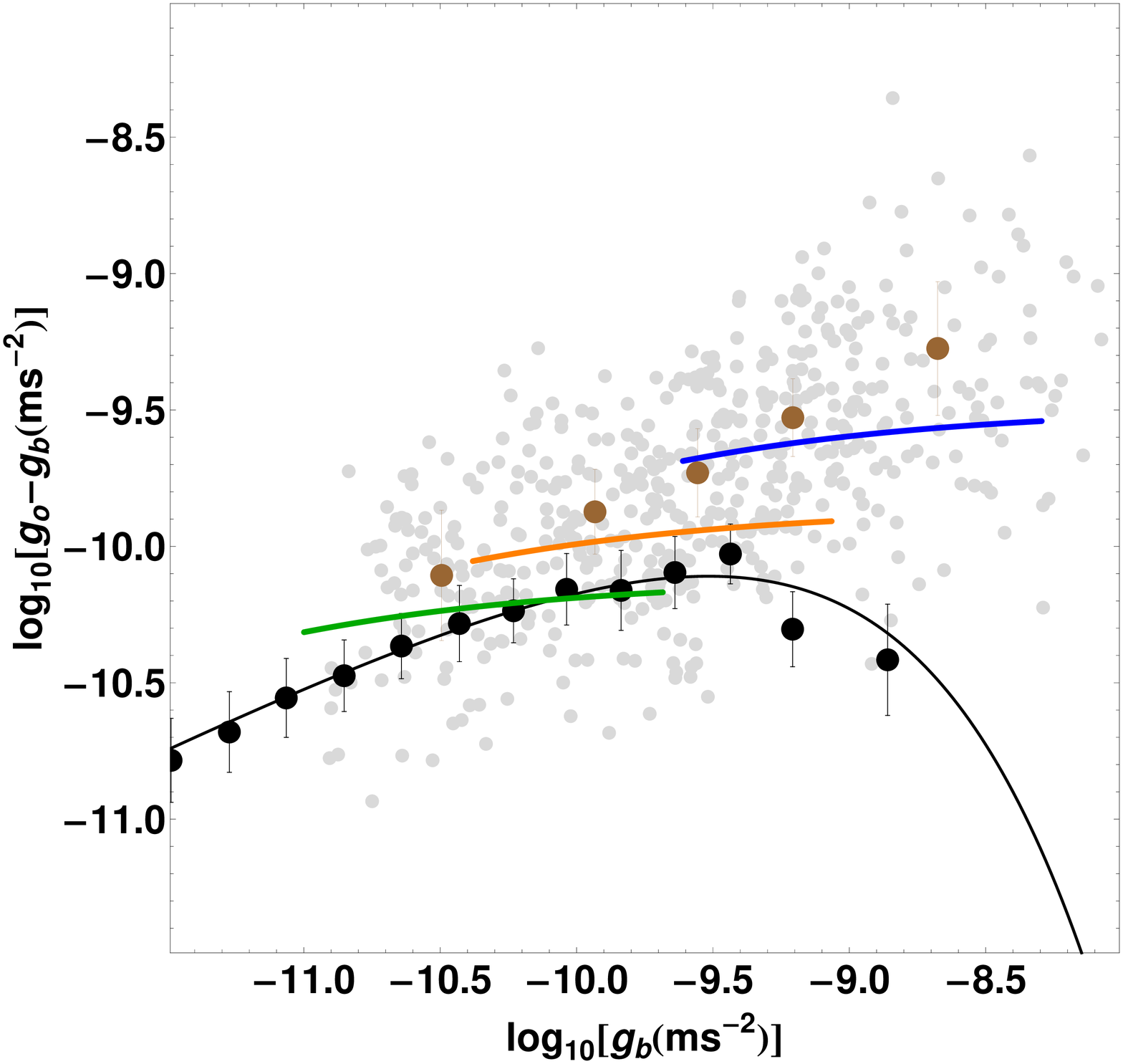}
        \includegraphics[width=0.9\columnwidth]{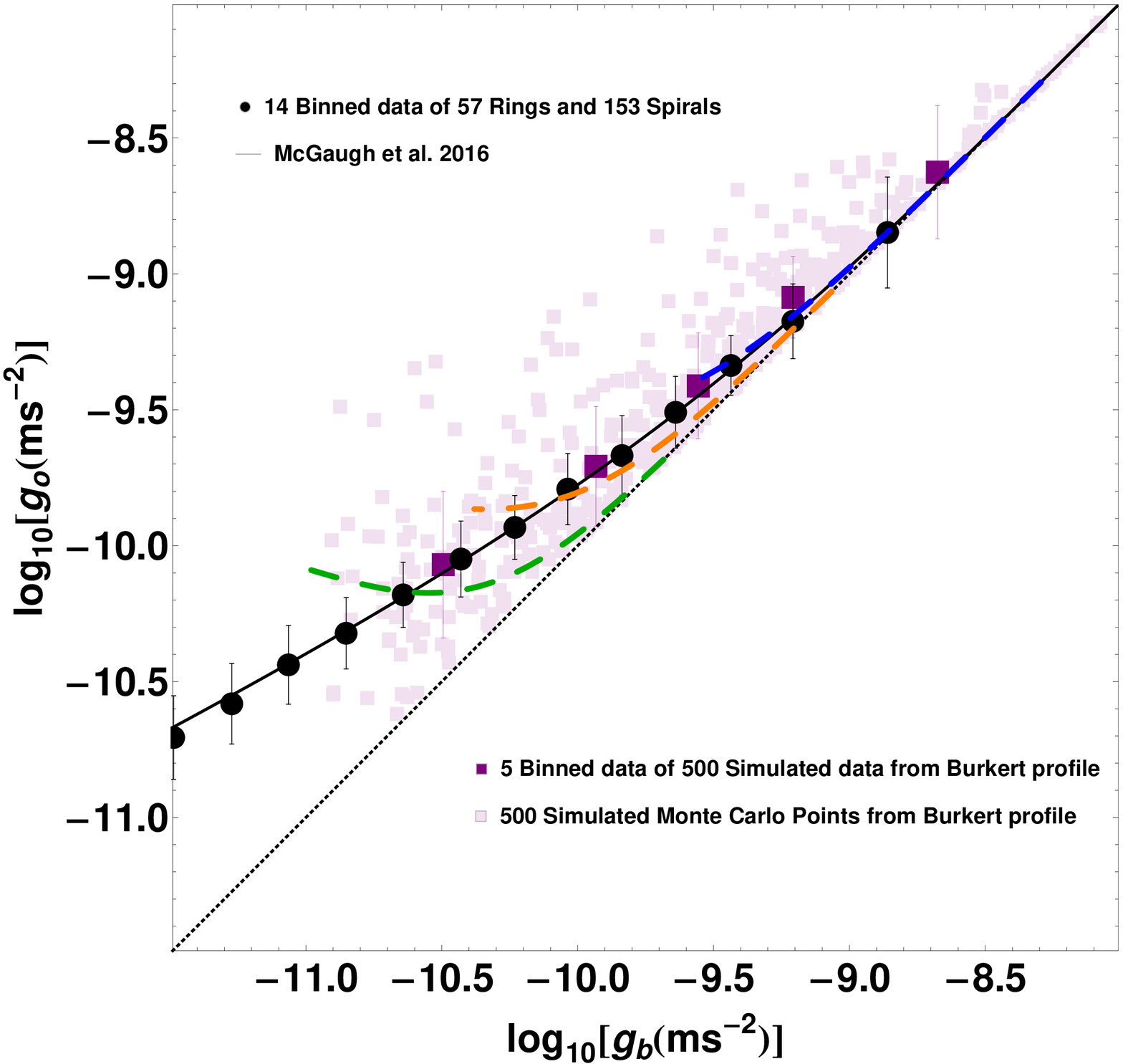}
        \includegraphics[width=0.9\columnwidth]{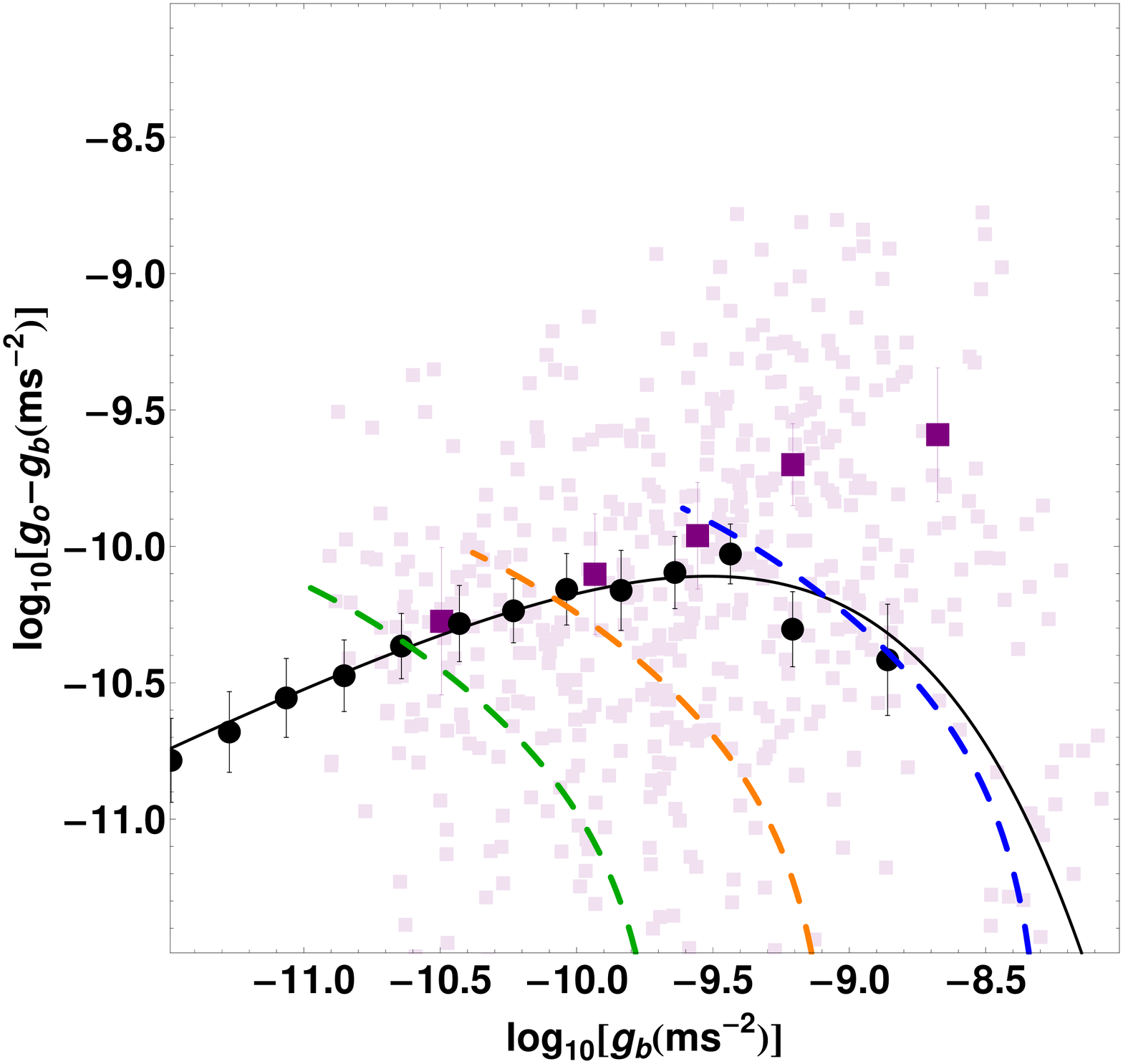}
        \caption{
        Semi-analytical model of two dark matter halo profiles.
        Upper row: NFW profile (solid line). Lower row: Burkert profile (dashed line).
        Left column: radial acceleration relation (RAR). Right column: halo acceleration relation (HAR).
        The black circles are binned data of all observational data from the sample
        (i.e., including both spiral and elliptical galaxies, see Fig.~\ref{fig:MHA}).
        The result of the Monte Carlo simulations is presented in grey circles for NFW profile (upper row),
        and light purple squares for Burkert profile (lower row).
        The brown circles and purple squares are the binned data of Monte Carlo results of NFW and Burkert profiles, respectively.
        }\label{fig:DM}
    \end{figure*}

    Under $\Lambda$CDM model, \citet{Navarro17} explained the RAR of spiral galaxies by a semi-analytical model of NFW profile \citep{Navarro96, Navarro97}.
    However, in an earlier study, \citet{McGaugh15} did not find clear pattern from NFW and baryonic profile.
    The key factor in \citet{Navarro17} is the adoption of an abundance matching relation of dark matter and baryonic mass \citep{Behroozi13}, which eliminated some mass discrepancy.

    Following the same procedure as \citet{Navarro17}, we work out the semi-analytical model of the dark matter halo of elliptical galaxies.
    We consider NFW and Burkert profiles, and three different galactic masses from inner region 0.01 to 2 effective radius ($R_{\rm e}$).
    The three mass paramters $M_{200}$:
    $5\times10^{11}$ $M_{\odot}$, $2\times10^{11}$ $M_{\odot}$ and $10^{11}$ $M_{\odot}$.
    The corresponding baryonic masses are estimated from the abundance matching \citep{Behroozi13}, and specifically,
    $1.1\times10^{10}$ $M_{\odot}$, $1.9\times10^{9}$ $M_{\odot}$ and $4.7\times10^{8}$ $M_{\odot}$, respectively.
    We set $r_{\rm s}=7R_{\rm e}$ \citep[][took $r_{\rm s}=5\times4/3R_{\rm e}$ for spiral galaxies]{Navarro17},
    and $c_{200}$: 30, 25, and 20 for NFW profile, and $r_{\rm s}=4R_{\rm e}$ and $c_{200}$: 6, 4 ,and 3 for Burkert profile.

    To assess the uncertainty in the model, Monte Carlo simulations are performed.
    Since different profiles or interpolating functions are more sensitive to the high acceleration region in the HAR,
    We work out 500 realizations for each profile of the mass parameter $M_{200}$ randomly picked from $10^{11} M_{\odot}$ to $10^{12} M_{\odot}$ with the corresponding baryonic mass by the abundance matching relation.
    We also consider the intrinsic scatters of $r_{\rm s}$ from $30\%$ Gaussian width of $c_{200}=25$ for NFW profile and $c_{200}=4$ for Burkert profile .

    The results are presented in Fig.~\ref{fig:DM}.
    In the figure, solid line and dashed line are semi-analytical model of NFW and Burkert, respectively.
    Grey circles and purple squares Monte Carlo results of NFW profile and Burkert, respectively.
    The brown circles and purple squares are the binned data of Monte Carlo results of NFW and Burkert profiles, respectively.
    The black circles are binned data of all observational data from the sample (i.e., including both spiral and elliptical galaxies, see Fig.~\ref{fig:MHA}).

    The HAR provides much rigorous test than the RAR even for the Burkert profile which is without cusp problem.
    As shown in the left row of Fig.~\ref{fig:DM}, both NFW profile (solid lines in upper panel) and Burkert profile (dash lines lower panel) are consistent with RAR from observation (black circles).
    However, both profiles have difficulty in HAR (right row of Fig.~\ref{fig:DM}.

    In semi-analytical model of NFW profile (upper-right panel), the halo acceleration $g_{\rm h}$ approaches a constant value at large baryonic acceleration $g_{\rm b}$ (corresponding to inner region and related to cusp problem of the profile).
    The binned data of Monte Carlo results of NFW (brown circles) also does not show the characteristic maximum of HAR.
    On the other hand, in semi-analytical model of Burkert profile (lower-right panel), $g_{\rm h}\rightarrow 0$ at large $g_{\rm b}$, but there is no clear maximum.
    However, its binned data of Monte Carlo results (purple squares)
    does not show a maximum in this range.

	As one may expect, the uncertainty in the mass deduced from Salpeter IMF would affect the result.
	We did $20\%$ uncertainty analysis in section 2.
    Indeed, the effect is noticeable in the high acceleration regime, in particular the last two binned data points (see Table \ref{tab:uncertainty}).
	However, the result of NFW and Burkert DM model on HAR still deviated a lot from the observed values and their uncertainty.
	HAR is a challenge to dark matter model.
	On the other hand, the parameters of different MOND interpolating functions will vary according to the uncertainty of Salpeter IMF.

    Our analysis on HAR shows that not only the data from low acceleration regime ($g_{\rm b}\ll \mathfrak{a}_0$), but also data from high acceleration regime ($g_{\rm b}\gg \mathfrak{a}_0$) is very important to the study of dark matter or modified gravity theory.
    The maximum of the HAR, in particular its position, may rule out many theories.
    To confirm the existence of such a maximum, data from high acceleration regime is highly desirable, such as inner region of high surface density galaxies.
    Moreover, strong gravitational lensing also can probe acceleration between $10^{-10}$ to $10^{-9}$ m\,s$^{-2}$.

    Finally, if there is a fundamental theory behind the interpolating function, the exact form will come from it.
    Here, we would like to stress on the method to test it for both modified law and dark matter model.

\section*{ACKNOWLEDGEMENTS}
    We are grateful to M. Milgrom for his perspective and comments and S. S. McGaugh for his suggestions.
    We appreciate the anonymous reviewer for valuable comments to improve the clarity of this work.
    This work is supported by the Taiwan Ministry of Science and Technology grant MOST 105-2112-M-008-011-MY3.

\end{document}